# Permanent-magnet-based transverse thermoelectric generator with high fill factor driven by anomalous Nernst effect

Fuyuki Ando,[1,a] Takamasa Hirai,[1] and Ken-ichi Uchida[1,a]

[1]*National Institute for Materials Science, Tsukuba 305-0047, Japan*
[a] Authors to whom correspondence should be addressed. ANDO.Fuyuki@nims.go.jp and UCHIDA.Kenichi@nims.go.jp

**ABSTRACT**

A transverse thermoelectric generator for magnetic-field-free and high-density power generation utilizing the anomalous Nernst effect is constructed and its performance is characterized. By alternately stacking two different permanent magnets with the large coercivity and anomalous Nernst coefficients of opposite sign, transverse thermoelectric voltage and power can be generated in the absence of external magnetic fields and enhanced owing to a thermopile structure without useless electrode layers. In the permanent-magnet-based stack, the magnetic attractive force enables easy construction of the thermopile structure with a high fill factor. In this study, we construct a bulk module consisting of twelve pairs of $SmCo_5$- and $Nd_2Fe_{14}B$-type permanent magnets with respectively having the positive and negative anomalous Nernst coefficients, whose fill factor reaches ~80%, whereas that of conventional thermoelectric modules based on the Seebeck effect is typically 30~60%. We demonstrate magnetic-field-free anomalous Nernst power generation up to 177 μW at a temperature difference of 75 K around room temperature, which corresponds to the largest anomalous Nernst power density of 65 μW/cm$^2$. The presented module structure concept will provide a design guideline for high-performance transverse thermoelectric power generation.

**MANUSCRIPT**

In the urgent issue of sustainable energy solutions, thermoelectric generation technology attracts much attention as a promising avenue for harvesting waste heat and converting it into electricity. The conventional thermoelectric generation is driven by the Seebeck effect, which induces an electric field **E** in the direction parallel to a temperature gradient ∇*T*. Thus, the Seebeck effect is classified into a longitudinal thermoelectric effect. Due to the parallel relationship between **E** and ∇*T*, a thermoelectric generator (TEG) based on the Seebeck effect needs to consist of many legs of *p*- and *n*-type semiconductors, which are arranged in isolation to prevent a short-circuit fault and connected with each other in series on the hot and cool sides by electrode materials to enhance the output power. Such a Π-shaped module configuration imposes two problems that reduce the output power. The first problem is that the temperature gradient and optimum current density in the thermoelectric materials respectively decrease due to the contact thermal and electrical resistances at the electrode junctions, which especially deteriorate faced on the hot side.[1–3] The second problem is caused by the low fill factor which is the area density of the active thermoelectric materials within the module. As the fill factor decreases, the power density proportionally decreases, and a part of the input heat flow is wasted as radiation and convection heat losses.[4–6] The fill factor of conventional bulk TEGs based on the Seebeck effect is typically in the range of 30~60%[5,7,8] and at most ~70%.[9] These two problems prevent the efficient power generation, resulting in limited applications of the thermoelectric generation technology.

A key to overcome the first problem is a utilization of transverse thermoelectric effects, which generate an electric field **E** in the direction perpendicular to $\nabla T$. Owing to the orthogonal relationship between **E** and $\nabla T$, the electric circuit of transverse TEGs can form without junctions faced on the hot side and scale up simply by elongating the length of thermoelectric materials along **E** without increasing the number of junctions.[10–15] To utilize this geometrical advantage, many studies have recently focused on developing physics, materials science, and applications of transverse thermoelectric phenomena.[3] Among the various transverse thermoelectric phenomena, studies on the anomalous Nernst effect (ANE) have rapidly progressed with the development of topological materials science and spin caloritronics.[16–22] As shown in Fig. 1(a), the ANE-driven electric field **E**$_{\mathrm{ANE}}$ is in the cross-product direction (*y*-axis) of $\nabla T$ (*x*-axis) and the spontaneous magnetization (*z*-axis) in a magnetic material as the following equation:

$$\mathbf{E}_{\mathrm{ANE}} = S_{\mathrm{ANE}}(\mathbf{m} \times \nabla T), \tag{1}$$

where $S_{\mathrm{ANE}}$ is the anomalous Nernst coefficient and **m** is the unit vector of the magnetization. Typical TEGs based on ANE or the ordinary Nernst effect have been proposed with the thermopile and coil geometries having the advantages of simplicity for electrical wiring and suppression of the thermal deterioration at the electrode junctions.[10–13]

However, conventional transverse TEGs have other problems. The aforementioned second problem regarding the fill factor still remains in conventional transverse TEGs. The scalable Nernst device with the coil geometry has been recently demonstrated with a relatively high fill factor (~70%) by thinning the insulator layers.[13] Meanwhile, the fill factor of transverse TEG with the thermopile geometry was still 26% because it requires the adequate spaces to form a zigzag circuit by the single thermoelectric materials and electrodes.[14] Therefore, the effective way to improve the fill factor in the thermopile geometry needs to be established. Furthermore, ANE and the ordinary Nernst effect in many materials work only under external magnetic fields.[10,11,22,23] In the case of ANE, the elongation of the magnetic materials in *y*-axis increases the ANE-induced thermoelectric voltage, whereas **m** necessarily directs along the hard axis of shape magnetic anisotropy (*z*-axis) due to the orthogonal relationship between **E**$_{\mathrm{ANE}}$ and **m**. To solve this issue, the use of permanent magnets such as $SmCo_5$- and $Nd_2Fe_{14}B$-type magnets has recently been proposed, which enables the magnetic-field-free operation of ANE owing to the large coercivity and remanent magnetization.[23–25] However, the use of permanent magnets in the thermopile geometry has not been demonstrated so far.

In this study, we present a concept of bulk transverse TEGs based on ANE to realize the high fill factor and magnetic-field-free operation simultaneously. Figures 1(a) and (b) show the schematics of the components and module structure for proposed transverse TEG, respectively. Transverse TEG consists of alternately stacked two different permanent magnets with positive and negative $S_{\mathrm{ANE}}$ intermediated by thin insulator layers. As the proof-of-concept demonstration, we fabricate permanent-magnet-based transverse TEG comprising many $SmCo_5$- and $Nd_2Fe_{14}B$-type permanent magnet slabs respectively having positive and negative $S_{\mathrm{ANE}}$.[23] The use of permanent magnets as the components in the thermopile structure allows us to easily construct TEG with a fill factor of ~80%, which is much higher than the fill factors in the literature.[13,14] All the $SmCo_5$ and $Nd_2Fe_{14}B$ slabs are confirmed to positively contribute to the total transverse thermoelectric output by the lock-in thermography (LIT) method.[23–28] An ANE-induced thermoelectric voltage is generated without an external magnetic field and its magnitude is ideally equal to the calculated one from $S_{\mathrm{ANE}}$ of $SmCo_5$ and $Nd_2Fe_{14}B$. As a result of the structural design, the maximum ANE-induced output power reaches 177 μW at a temperature difference $\Delta T = 75$ K, which corresponds to the ANE-induced power density of 65 μW/cm$^2$. This power density is the largest value among transverse TEGs based on ANE.[29]

The important point of transverse TEG structure in Fig. 1(b) is to utilize not only the positive and negative $S_{ANE}$ values but also the magnetic attractive force due to the remanent magnetization of the permanent magnets. The series circuit can be formed by inserting the thin insulator layers between the neighboring magnets and electrically connecting the ends of magnets to sum up the transverse thermoelectric voltage in a zigzag way. The circuit structure drastically increases the fill factor by the reduction of superfluous spaces for electrical wiring between the magnets. The magnetic attractive forces between the permanent magnets allow the construction of a dense stack of themselves and stabilization of the remanent magnetization along $z$-axis owing to the stray field of neighboring magnets.

First, we investigated the thermoelectric and magnetic properties of the $SmCo_5$- and $Nd_2Fe_{14}B$-type permanent magnet slabs. We used $SmCo_5$ and $Nd_2Fe_{14}B$ circular disks with a thickness of 0.5 mm and diameter of 20 mm having the easy axis of magnetic anisotropy along the thickness direction, which were commercially available from Magfine Corporation. The surface of the $Nd_2Fe_{14}B$ disks was coated by a Ni-Cu-Ni plating layer with a thickness of ~15 μm to prevent oxidization.[30] To quantitatively estimate $S_{ANE}$ of these magnets without applying an external magnetic field, we used the LIT method. The LIT method enables the measurements of the anomalous Ettingshausen effect (AEE), the reciprocal effect of ANE, and the estimation of $S_{ANE}$ through the Onsager reciprocal relation: $\Pi_{AEE} = S_{ANE}T$ with $\Pi_{AEE}$ being the anomalous Ettingshausen coefficient. The experimental procedure to estimate $\Pi_{AEE}$ is the same as that in the previous reports.[23–28] Figure 2(a) shows the measurement results of $S_{ANE}$ and $\Pi_{AEE}$ at 300 K in the absence of external magnetic fields. The $S_{ANE}$ values for $SmCo_5$ and $Nd_2Fe_{14}B$ in the remanent states were estimated to be $+3.5 \times 10^{-6}$ and $-8.7 \times 10^{-7}$ V/K, respectively, the magnitude and sign of which are consistent with those reported in Ref. 23. Figures 2(b) and (c) respectively show the electrical conductivity $\sigma$ measured by a four-probe method and the thermal conductivity $\kappa$ estimated from thermal diffusivity measurements using the laser flash analyzer and differential scanning calorimetry. According to the obtained $S_{ANE}$, $\sigma$, and $\kappa$ values, the transverse thermoelectric figure of merit $z_{ANE}T$ ($= S_{ANE}^2 \sigma T / \kappa$) for $SmCo_5$ and $Nd_2Fe_{14}B$ were estimated to be $4.2 \times 10^{-4}$ and $2.5 \times 10^{-5}$ at 300 K, respectively. Meanwhile, we measured the magnetization $M$ curves for these magnets with applying an external magnetic field **H** along the thickness direction to confirm the remanent magnetization. Figure 2(d) shows the results of the $M$-$H$ curves at room temperature. Both the $SmCo_5$ and $Nd_2Fe_{14}B$ slabs show the remanent magnetization comparable to the saturation magnetization and the large coercivity, enabling the magnetic-field-free operation of ANE.

The next step is to fabricate high-density TEG using the pairs of the $SmCo_5$ and $Nd_2Fe_{14}B$ slabs. Figure 2(e) shows the photograph of our transverse TEG consisting of alternately stacked twelve $SmCo_5$/$Nd_2Fe_{14}B$ pairs. We inserted ~0.05-mm-thick paper towel layers soaked with heat-resistant glue (Aron Alpha Tough-power) between the slabs to avoid the reduction of the ANE voltage due to shunting effects. After the glue cured, we cut the stacked discs into a rectangular shape with a size of 8.2 mm ($\| \nabla T$) × 16.5 mm ($\|$ **E**) × 16.5 mm ($\|$ the stacking direction) so that the $Nd_2Fe_{14}B$ surfaces were exposed on the planes attaching to electrodes and heat source/sink by the removal of the Ni-Cu-Ni plating layers. The ends of the neighboring $SmCo_5$ and $Nd_2Fe_{14}B$ slabs were manually and carefully connected by attaching indium to form a zigzag series circuit without a short-circuit fault. Then, the stacked $SmCo_5$ and $Nd_2Fe_{14}B$ slabs were magnetized by applying a pulse magnetic field of +8 T in the stacking direction. The adhesion of the slabs is reinforced by their magnetic attractive force. In this structure, the ANE voltage in all the slabs positively contributes to the total voltage owing to the positive (negative) $S_{ANE}$ of $SmCo_5$ ($Nd_2Fe_{14}B$) when the $\nabla T$ and **m** directions in all the slabs are the same. The inset of Fig. 2(e) shows a top view of TEG, which confirms that the slabs are stacked with high density and separated by the thin insulator layers.

To characterize the fill factor, we observed the infrared image of the top surface of $SmCo_5$/$Nd_2Fe_{14}B$-based TEG [Fig. 2(f)]. The fill factor can be estimated from the contrast of thermally emitted infrared intensity due to the difference in infrared emissivity between the metallic surfaces with low emissivity and paper towels with high emissivity. Here, the line profiles are taken along $z$-axis as shown in Fig. 2(g), where the infrared intensity at each $z$-position is averaged along $y$-axis. The contrast of infrared intensity was clearly observed between the slabs and their spaces, and the relative area of the slabs was estimated to be 81%, when the threshold infrared intensity value is set to be 95 [Fig. 2(g)].

The contribution of the transverse thermoelectric conversion in each slab and the absence of a short-circuit fault in $SmCo_5$/$Nd_2Fe_{14}B$-based TEG is confirmed through the AEE measurements by the LIT technique. Figure 3(a) shows a schematic of the LIT measurement setup for TEG. To quantitatively estimate the surface temperature, we coated the top surface of TEG by an insulating black ink with high infrared emissivity (>0.94). The LIT measurements were performed while applying a square wave charge current $\mathbf{J_c}$ with the amplitude $J_c$, frequency $f$, and zero-offset.[23–28] The application of $\mathbf{J_c}$ generates an AEE-induced heat flow $\mathbf{J}_{Q,AEE}$ in the cross-product direction of $\mathbf{J_c}$ and $\mathbf{m}$ as[31]

$$\mathbf{J}_{Q,AEE} = \Pi_{AEE}(\mathbf{J_c} \times \mathbf{m}). \tag{2}$$

As a result of the $\mathbf{J}_{Q,AEE}$ generation, the temperature on the surfaces of the magnetic materials is modulated. By extracting the transient first-harmonic temperature modulation signals of the thermal images and transforming them into lock-in phase $\varphi$ and amplitude $A$ by Fourier analysis, the pure contribution of thermoelectric effects, such as AEE and the Peltier effect, can be visualized without the contamination by Joule heating.[26,27] In magnetized permanent magnets, the AEE-induced temperature modulation free from the Peltier effect can be measured in the absence of an external magnetic field in the areas far from the sample edges.[23] The spatial distribution of the temperature modulation allows us to confirm that all the $SmCo_5$ and $Nd_2Fe_{14}B$ slabs correctly operates without a short-circuit fault. Figures 3(b) and (c) show the examples of the $\varphi$ and $A$ images at $f$ = 1.0 Hz and $J_c$ = 1 A, respectively. The thermoelectric signals on the top surface of $SmCo_5$/$Nd_2Fe_{14}B$-based TEG were clearly observed with high area density. Importantly, the $SmCo_5$ and $Nd_2Fe_{14}B$ respectively show the spatially homogeneous $A$ signals, which is consistent with the feature of AEE with $\mathbf{m}$ along $z$-axis [Fig. 3(c)]. This uniformity confirms that $\mathbf{J_c}$ flows in a zigzag way [indicated by black arrows in Fig. 3(b)] without a short-circuit fault. Because of the opposite direction of $\mathbf{J_c}$ and opposite sign of $\Pi_{AEE}$ between $SmCo_5$ and $Nd_2Fe_{14}B$, the resultant temperature modulation due to AEE shows the same sign in all the slabs, which was confirmed by the almost same $\varphi$ values on the whole area of the top surface [Fig. 3(b)]. As shown in Fig. 3(c), the $SmCo_5$ slabs show the larger $A$ signals than the $Nd_2Fe_{14}B$ slabs due to larger $\Pi_{AEE}$ [Fig. 2(a)]. Figures 3(d) and (e) show the line profiles of $\varphi$ and $A$ taken along $z$-axis, where the signals are averaged along $y$-axis, within the area indicated by a 4.5 mm square in Figs. 3(b) and (c), respectively. The $Nd_2Fe_{14}B$ slabs show the large $\varphi$ delay and the rapid drop in $A$ as $f$ increases, which can be explained by the fact that $Nd_2Fe_{14}B$ has lower $\kappa$ than $SmCo_5$ [Fig. 2(c)].[27] Thus, the LIT measurements reveal that all the $SmCo_5$ and $Nd_2Fe_{14}B$ slabs positively contribute to the transverse thermoelectric conversion without a short-circuit fault, while the superfluous space for electrical wiring between the neighboring slabs are significantly reduced.

We are in a position to demonstrate magnetic-field-free transverse thermoelectric power generation in $SmCo_5$/$Nd_2Fe_{14}B$-based TEG. Figure 4(a) shows the schematic and photograph of the setup for four-terminal measurements of thermoelectric power, where TEG was sandwiched by a heater and heat sink. When TEG is practically used, the current source will be replaced to a load resistor so that one can obtain the output power without an external power supply. The temperature of the heat sink was controlled by flowing coolant at 273 K. The actual $\Delta T$ inside TEG was estimated by measuring the thermal image of its side surface coated by the black ink. The hot and cool side temperatures ($T_h$ and $T_c$) are defined as the highest and lowest temperatures of the thermal image in TEG along the heat flow direction.

To investigate the pure contribution of the ANE-induced thermopower independent of the parasitic Seebeck-effect-induced thermopower, we measured the dependence of the open circuit voltage $V_{oc}$ on the remanent magnetization direction. We repetitively reversed the remanent magnetization between the +z and -z directions, represented by +$M$ and -$M$ states, by applying a pulse magnetic field of 8 T and measured $V_{oc}$ as a function of $\Delta T$ in the absence of an external magnetic field for two cycles [Fig. 4(b)]. We found that the sign of $V_{oc}/\Delta T$ was reversed by reversing the magnetization direction as shown in the inset of Fig. 4(b), indicating that the $V_{oc}$ signal originates from pure ANE free from the offset due to the Seebeck effect. We confirmed that the $V_{oc}$ value is quantitatively consistent with the calculated value from $S_{ANE}$ of $SmCo_5$ and $Nd_2Fe_{14}B$ [Fig. 2(a)] as the following equation:

$$V_{oc} = \frac{nl[\tilde{S}_{ANE}(SmCo_5) - \tilde{S}_{ANE}(Nd_2Fe_{14}B)] \cdot \mathbf{m}}{h} \cdot \Delta T, \qquad (3)$$

where $n$ is the number of the $SmCo_5/Nd_2Fe_{14}B$ pairs, $l$ the length of each slab along the $\mathbf{E}_{ANE}$ direction, and $h$ the height of each slab along the $\nabla T$ direction, and $\tilde{S}_{ANE}$ the averaged $S_{ANE}$ value at temperatures ranging from $T_c$ to $T_h$. Here, we approximated $\tilde{S}_{ANE}$ for $SmCo_5$ and $Nd_2Fe_{14}B$ by the $S_{ANE}$ values at room temperature. Figure 4(b) shows the good matching between the measured and calculated $V_{oc}$ values, which confirms the fact that the thermoelectric voltage in our TEG is due purely to ANE.

The magnetic-field-free operation of the ANE-induced power generation was demonstrated by measuring the thermoelectric voltage $V$ with applying a load current $I_{load}$ to TEG. Figure 4(c) shows the $I_{load}$ dependence of the output power $P$ at various values of $\Delta T$, where $P$ is the product of $V$ and $I_{load}$. The high $P$, 177 μW at maximum when $\Delta T = 75$ K and $I_{load} = 41$ mA, was obtained owing to both the large $V_{oc}$ and small internal resistance of module $R_{module}$, which is the slope of the $I_{load}$-$V$ curve. The $R_{module}$ of TEG was estimated to be 9.9 mΩ at $\Delta T = 21$ K, whereas the value calculated from the $\sigma$ values and dimensions of $SmCo_5$ and $Nd_2Fe_{14}B$ is 9.1 mΩ; the deviation of $R_{module}$ from the calculated value due to the electrodes and junctions is only +8%. The inset of Fig. 4(c) shows the $\Delta T$ dependences of the maximum output power $P_{max}$ and power density per unit area $\omega_{max}$, which is proportional to the fill factor. Owing to the high $P_{max}$ of 177 μW and high fill factor of ~80%, the resultant $\omega_{max}$ reaches 65 μW/cm$^2$ at $\Delta T = 75$ K, which is the highest value among transverse TEGs based on ANE.[29]

Finally, we discuss future developments and applications of permanent-magnet-based transverse TEG. Since the fill factor of TEG is determined by the thickness ratio between the magnetic materials and insulator layers, we can further increase the fill factor and $\omega_{max}$ by using thinner insulator layers. Although the demonstrated $P_{max}$ of 177 μW can be used, for example, as stand-alone power supply for wireless sensor network systems, the further improvement of $z_{ANE}T$ in permanent magnets is required for widespread thermoelectric applications. This is because the commercial TEGs based on the Seebeck effect exhibit 70~80 mW/cm$^2$ at temperature differences of 60~70 K around room temperature[32,33], which are still three orders of magnitude larger than that of the presented transverse TEG based on ANE. Decreasing $\kappa$ and increasing $S_{ANE}$ are the significant tasks, especially for permanent magnets with negative $S_{ANE}$, since small negative $S_{ANE}$ of $Nd_2Fe_{14}B$ is a bottleneck in the output of our TEG. Additionally, the thermal durability of TEG is limited by that of the remanent magnetization in $Nd_2Fe_{14}B$-type permanent magnets, which gradually decreases beyond ~400 K.[34] Meanwhile, since the magnetic force of $SmCo_5$- and $Nd_2Fe_{14}B$-type permanent magnets is at the highest level, our TEG enables energy harvesting from everywhere permanent magnets are used.

In conclusion, we constructed ANE-based transverse TEG with the high fill factor and demonstrated its magnetic-field-free operation. As the proof-of-concept demonstration, we used the SmCo$_5$- and Nd$_2$Fe$_{14}$B-type permanent magnets with the large coercivities, large remanent magnetization, and $S_{\text{ANE}}$ of the opposite sign. The pairs of the SmCo$_5$ and Nd$_2$Fe$_{14}$B slabs were integrated into high-density TEG, whose fill factor surprisingly reaches ~80%. Using the LIT method, all the SmCo$_5$ and Nd$_2$Fe$_{14}$B slabs were confirmed to positively contribute to the transverse thermoelectric conversion without a short-circuit fault. An ANE-induced output power of 177 μW was generated at a temperature difference of 75 K without an external magnetic field, which corresponds to the largest power density of 65 μW/cm$^2$ among TEGs based on ANE. This study manifests the importance to develop permanent magnet materials with large positive and negative $S_{\text{ANE}}$ values together with large remanent magnetization. The device architecture proposed here will be a design guideline for high-performance transverse thermoelectric power generation.

The authors thank H. Sepehri-Amin and Y. Sakuraba for valuable discussions and K. Suzuki and M. Isomura for technical supports. This work was supported by ERATO "Magnetic Thermal Management Materials" (No. JPMJER2201) from JST, Japan, and NEC Corporation.

## AUTHOR DECLARATIONS

**Conflict of Interest**

The authors have no conflicts to disclose.

## DATA AVAILABILITY

The data that support the findings of this study are openly available in Zenodo at https://doi.org/10.5281/zenodo.10005678.

## REFERENCES

[1] G. Min, and D.M. Rowe, J. Power Sources **38**, 253–259 (1992).

[2] S. Shittu, G. Li, X. Zhao, and X. Ma, Appl. Energy **268**, 115075 (2020).

[3] K. Uchida, and J.P. Heremans, Joule **6**, 2240–2245 (2022).

[4] M. Gomez, R. Reid, B. Ohara, and H. Lee, J. Appl. Phys. **113**, 174908 (2013).

[5] P. Ying, H. Reith, K. Nielsch, and R. He, Small **18**, 2201183 (2022).

[6] F. Tohidi, S. Ghazanfari Holagh, and A. Chitsaz, Appl. Therm. Eng. **201**, 117793 (2022).

[7] Z. Bu, X. Zhang, Y. Hu, Z. Chen, S. Lin, W. Li, and Y. Pei, Energy Environ. Sci. **14**, 6506–6513 (2021).

[8] C. Xu, Z. Liang, W. Ren, S. Song, F. Zhang, and Z. Ren, Adv. Energy Mater. **12**, 2202392 (2022).

[9] F. Ando, H. Tamaki, Y. Matsumura, T. Urata, T. Kawabe, R. Yamamura, Y. Kaneko, R. Funahashi, and T. Kanno, Mater. Today Phys. **36**, 101156 (2023).

[10] M.H. Norwood, J. Appl. Phys. **34**, 594–599 (1963).

[11] Y. Sakuraba, Scr. Mater. **111**, 29–32 (2016).

[12] M. Ikhlas, T. Tomita, T. Koretsune, M.T. Suzuki, D. Nishio-Hamane, R. Arita, Y. Otani, and S. Nakatsuji, Nat. Phys. **13**, 1085–1090 (2017).

[13] Z. Yang, E.A. Codecido, J. Marquez, Y. Zheng, J.P. Heremans, and R.C. Myers, AIP Adv. **7,** 095017 (2017).

[14] M. Murata, K. Nagase, K. Aoyama, A. Yamamoto, and Y. Sakuraba, iScience **24**, 101967 (2021).

[15] M.R. Scudder, B. He, Y. Wang, A. Rai, D.G. Cahill, W. Windl, J.P. Heremans, and J.E. Goldberger, Energy Environ. Sci. **14**, 4009–4017 (2021).

[16] K. Uchida, H. Adachi, T. Kikkawa, A. Kirihara, M. Ishida, S. Yorozu, S. Maekawa, and E. Saitoh, Proc. IEEE **104**, 1946–1973 (2016).

[17] S.R. Boona, K. Vandaele, I.N. Boona, D.W. McComb, and J.P. Heremans, Nat. Commun. **7**, 13714 (2016).

[18] A. Sakai, Y.P. Mizuta, A.A. Nugroho, R. Sihombing, T. Koretsune, M.T. Suzuki, N. Takemori, R. Ishii, D. Nishio-Hamane, R. Arita, P. Goswami, and S. Nakatsuji, Nat. Phys. **14**, 1119–1124 (2018).

[19] H. Reichlova, R. Schlitz, S. Beckert, P. Swekis, A. Markou, Y.C. Chen, D. Kriegner, S. Fabretti, G. Hyeon Park, A. Niemann, S. Sudheendra, A. Thomas, K. Nielsch, C. Felser, and S.T.B. Goennenwein, Appl. Phys. Lett. **113**, 212405 (2018).

[20] W. Zhou, K. Yamamoto, A. Miura, R. Iguchi, Y. Miura, K. Uchida, and Y. Sakuraba, Nat. Mater. **20**, 463–467 (2021).

[21] K. Uchida, W. Zhou, and Y. Sakuraba, Appl. Phys. Lett. **118**, 140504 (2021).

[22] A. Von Ettingshausen, and W. Nernst, Ann. Phys. **265**, 343 (1886).

[23] A. Miura, H. Sepehri-Amin, K. Masuda, H. Tsuchiura, Y. Miura, R. Iguchi, Y. Sakuraba, J. Shiomi, K. Hono, and K. Uchida, Appl. Phys. Lett. **115**, 222403 (2019).

[24] A. Miura, K. Masuda, T. Hirai, R. Iguchi, T. Seki, Y. Miura, H. Tsuchiura, K. Takanashi, and K. Uchida, Appl. Phys. Lett. **117**, 082408 (2020).

[25] R. Modak, Y. Sakuraba, T. Hirai, T. Yagi, H. Sepehri-Amin, W. Zhou, H. Masuda, T. Seki, K. Takanashi, T. Ohkubo, and K. Uchida, Sci. Technol. Adv. Mater. **23**, 767–782 (2022).

[26] T. Seki, R. Iguchi, K. Takanashi, and K. Uchida, Appl. Phys. Lett. **112**, 152403 (2018).

[27] R. Das, R. Iguchi, and K. Uchida, Phys. Rev. Appl. **11**, 034022 (2019).

[28] K. Uchida, S. Daimon, R. Iguchi, and E. Saitoh, Nature **558**, 95–99 (2018).

[29] G. Lopez-Polin, H. Aramberri, J. Marques-Marchan, B.I. Weintrub, K.I. Bolotin, J.I. Cerdá, and A. Asenjo, ACS Appl. Energy Mater. **5**, 11835–11843 (2022).

[30] K. Uchida, T. Hirai, F. Ando, and H. Sepehri-Amin, Adv. Energy Mater. 2302375 (2023).

[31] T. Seki, R. Iguchi, K. Takanashi, and K. Uchida, J. Phys. D: Appl. Phys. **51**, 254001 (2018).

[32] KELK Ltd., Products information on thermo generation module, https://www.kelk.co.jp/english/generation/index.html.

[33] Coherent Corp., Thermoelectric Generator (TEG) Modules, https://ii-vi.com/product/thermoelectric-generator-teg-modules/.

[34] D. Brown, B.-M. Ma, and Z. Chen, J. Magn. Magn. Mater. **248**, 432–440 (2002).

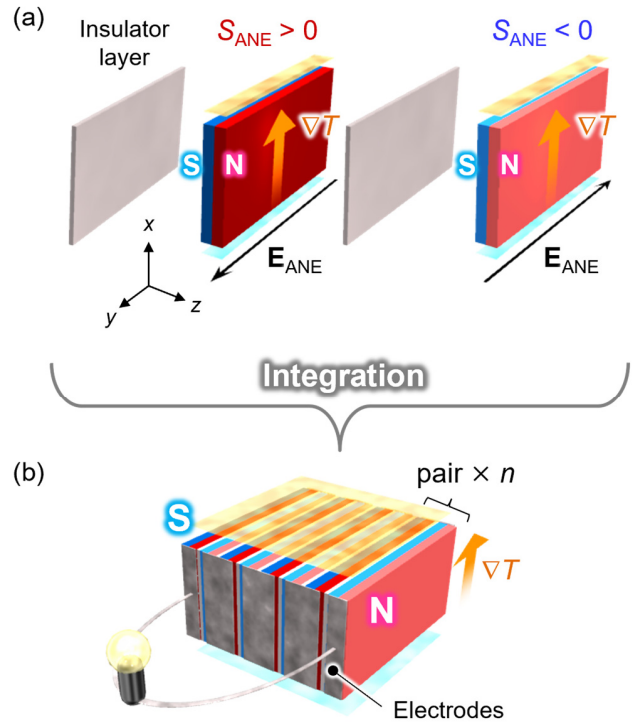

FIG. 1. Schematics of permanent magnets showing ANE and insulator layers (a) and integrated transverse TEG (b). The electric field induced by ANE $\mathbf{E}_{ANE}$ is generated in the cross-product direction ($y$-axis) of a temperature gradient $\nabla T$ ($x$-axis) and the spontaneous magnetization ($z$-axis) in a magnetic material. By alternately stacking two different permanent magnets with positive and negative anomalous Nernst coefficients $S_{ANE}$ intermediated by insulator layers, a high-density module is obtained. $n$ is the number of the pairs of the permanent magnets. Electrodes are attached at the ends of the permanent magnets.

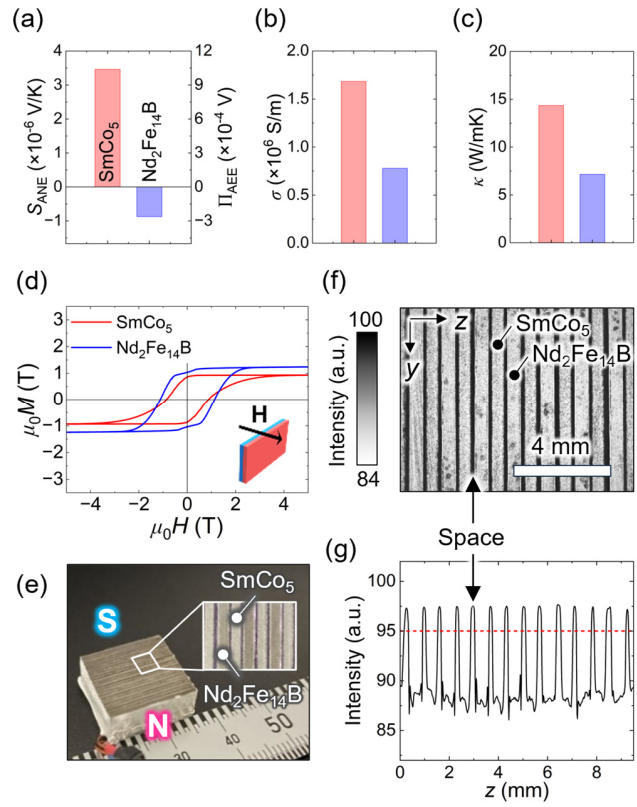

FIG. 2. Measurement results of the anomalous Nernst coefficient $S_{\text{ANE}}$ (a), electrical conductivity $\sigma$ (b), thermal conductivity $\kappa$ (c), and magnetization $M$ curves (d) for the SmCo$_5$- and Nd$_2$Fe$_{14}$B-type permanent magnets. (e) Photograph of transverse TEG composed of twelve pairs of the SmCo$_5$ and Nd$_2$Fe$_{14}$B slabs. The inset shows the top surface of TEG. (f) The infrared image of the top surface. (g) Line profile of the infrared intensity along $z$-axis.

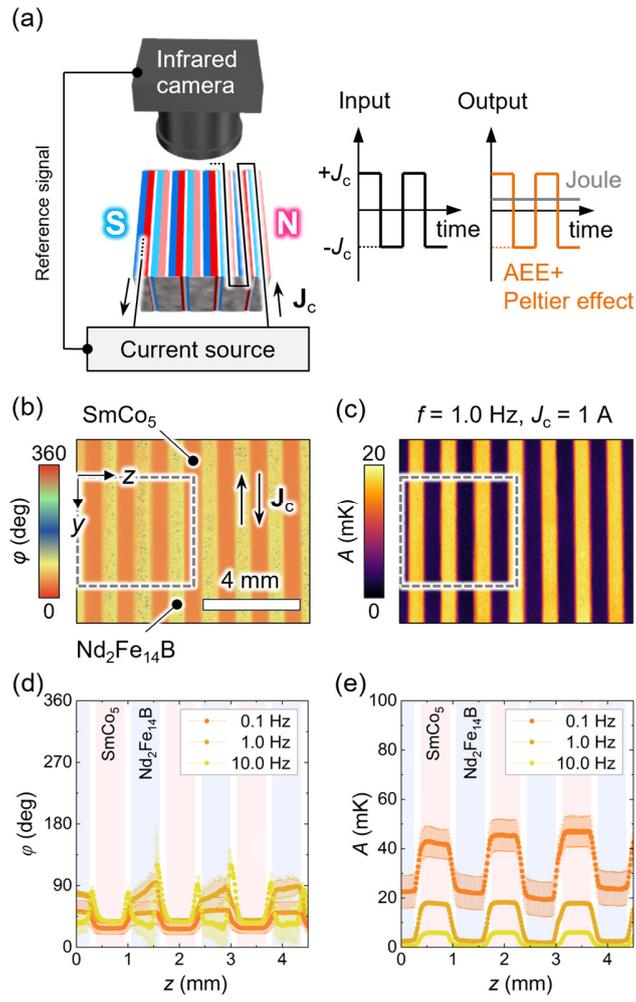

FIG. 3. (a) Schematic of the LIT measurement. The input charge current $\mathbf{J}_c$ flows in transverse TEG in a zigzag way. The temporal thermal image of the top surface is measured to visualize pure contributions of thermoelectric effects. The lock-in phase $\varphi$ (b) and amplitude $A$ (c) images at $f = 1.0$ Hz and $J_c = 1$ A. Line profiles of $\varphi$ (d) and $A$ (e) along $z$-axis in the dotted area of (b) and (c), respectively, for various values of $f$.

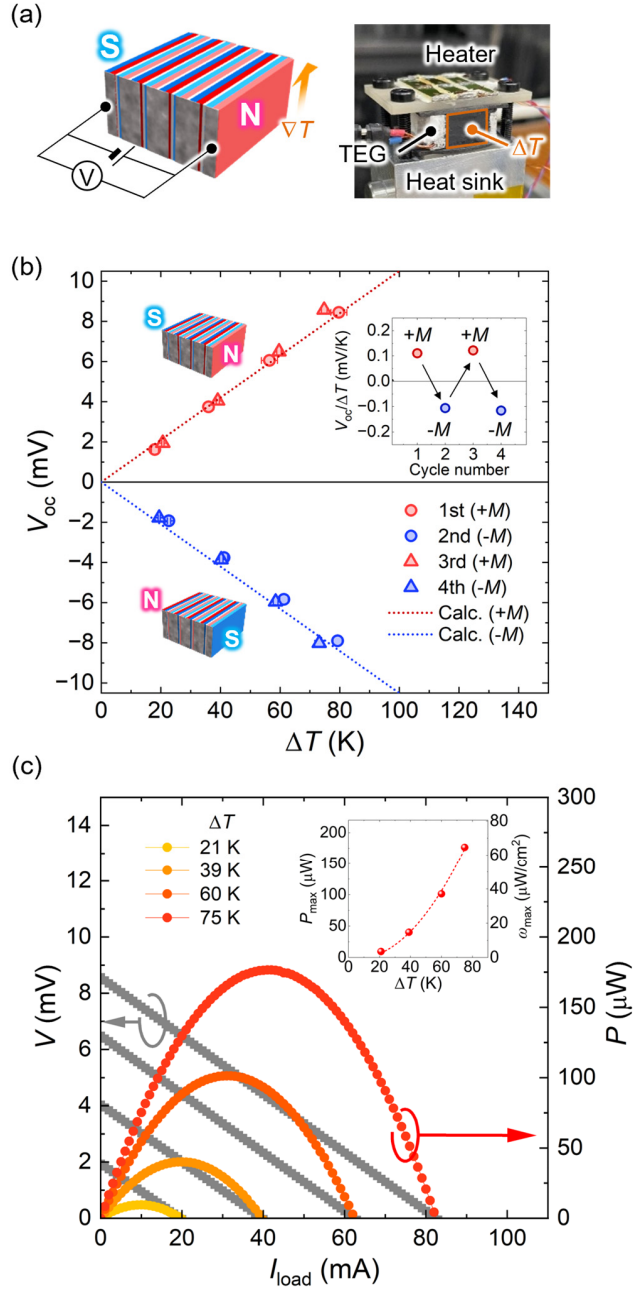

FIG. 4. (a) Schematic and photograph of transverse TEG to show the setup for four-terminal measurements of thermoelectric power and the area for estimating the temperature difference $\Delta T$ by a thermography. (b) $\Delta T$ dependence of the open circuit voltage $V_{oc}$ when the remanent magnetization is along the $+z$ direction ($+M$) and $-z$ direction ($-M$). (c) Load current $I_{load}$ dependence of the thermoelectric voltage $V$ and output power $P$ in the $+M$ state at various values of $\Delta T$. The inset shows the $\Delta T$ dependences of the maximum output power $P_{max}$ and power density per unit area $\omega_{max}$.

11